\documentclass[iop]{emulateapj}
\slugcomment{{\sc Accepted to ApJ:} January 16, 2012} 
\usepackage{natbib}
\usepackage{graphicx}
\usepackage{color}
\usepackage[normalem]{ulem}
\usepackage{amsmath}

\slugcomment{Not to appear in Nonlearned J., 45.}

\shorttitle{A CME\,--\,CME Interaction, and its radio and white-light manifestations}
\shortauthors{Martinez Oliveros et al.}

\begin{document}


\title{\textbf{\sc{The 2010 August 01 type II burst: A CME\,--\,CME Interaction, and its radio and white-light manifestations}}}


\author{\normalsize{\sc{Juan Carlos Mart\'inez Oliveros\altaffilmark{1}, Claire L. Raftery\altaffilmark{1}, Hazel M. Bain\altaffilmark{1},\\ Ying Liu\altaffilmark{1}, Vratislav Krupar\altaffilmark{2,3},
Stuart Bale\altaffilmark{1,4} and S\"am Krucker\altaffilmark{1,5}}}}


\altaffiltext{1}{Space Sciences Laboratory, University of California, Berkeley, USA}
\altaffiltext{2}{Institute of Atmospheric Physics, Academy of Sciences of the Czech Republic, Prague, Czech Republic}
\altaffiltext{3}{Faculty of Mathematics and Physics, Charles University, Prague, Czech Republic}
\altaffiltext{4}{Physics Department, University of California, Berkeley, USA}
\altaffiltext{5}{Institute of 4D Technologies, School of Engineering, University of Applied Sciences North Western Switzerland, 5210 Windisch, Switzerland}


\begin{abstract}

We present observational results of a type II burst associated with a CME\,--\,CME interaction observed in the radio and white-light wavelength range.  We applied radio direction-finding techniques to observations from the \textit{STEREO} and \textit{Wind} spacecraft, the results of which were interpreted using white-light coronagraphic measurements for context. The results of the multiple radio-direction finding techniques applied were found to be consistent both with each other and with those derived from the white-light observations of coronal mass ejections (CMEs).The results suggest that the Type II burst radio emission is causally related to the CMEs interaction.

\end{abstract}





\section{Introduction}

Over periods of increased solar activity, several coronal mass ejections (CMEs) can be launched by the same or nearby active regions \citep{2005IAUS..226..367G}. During these times of high activity, one or more of these CMEs may interact while propagating through the interplanetary medium. Almost a decade after the Þrst observations of  \emph{CME\,--\,associated} shock  \emph{regions} \citep{1987JGR....92.5725B}, CME\,--\,CME interactions were observed, at long wavelengths and in white-light coronagraphic images, by \citet{2001ApJ...548L..91G,2002ApJ...572L.103G} and \citet{2004P&SS...52.1399G}. The radio observations were obtained by the Radio and Plasma Wave Experiment \citep[WAVES,][]{1995SSRv...71..231B} on board the \textit{Wind} spacecraft, while the white-light observations were obtained by the Large Angle and Spectroscopic Coronagraph \citep{lasco} on board the \textit{Solar and Heliospheric Observatory} mission. Based on the observational characteristics of the CMEs from white-light coronagraph and radio observations, \citet{2004P&SS...52.1399G} concluded that the type II radio emission is enhanced and modified due to the interaction between two CMEs. \citet{2001ApJ...548L..91G} suggested that the observed radio enhancements result from the increased density in the upstream medium that reduces the Alfv\'en speed, thereby increasing the Mach number of the shock. This is in agreement with results from numerical simulations, confirming that the radio enhancement was likely to be produced at the interaction region shock \citep[e.g.,][]{2004A&A...415..755V}. \citet{2001ApJ...548L..91G} also mentioned additional possibilities for electron acceleration, such as reconnection between the two CMEs \citep[see also][]{2004P&SS...52.1399G}.

White-light (WL) imagers such as those on board the \textit{Solar TErrestrial RElations Observatory} \citep[\textit{STEREO};][]{Kaiser08} allow us to observe CMEs out to $\sim$1~AU with the Sun\,--\,Earth Connection Coronal and Heliospheric Investigation \citep[SECCHI;][]{Howard08}. The Cor1 and Cor2 coronagraphs, along with the heliospheric imagers (HI1 and HI2) on board, observe CMEs at visible wavelengths as they propagate through the heliosphere. Coronograph observations typically reveal the three part configuration of CMEs: the bright, dense core that is thought to be the erupting filament; the dark, low density cavity surrounding the core; and the bright front, or the leading edge.  

The  stereoscopic observations of the \textit{STEREO} mission allow us to determine the location of different CME features in three dimensions  \citep[e.g.,][]{{2009ApJ...695..636F},{Aschwanden2010},{2009ApJ...691L.151L},{2010ApJ...710L..82L},{2010ApJ...712..453M}}.  A similar approach can be used to determine the position of radio sources in what is called radio direction-finding. Several direction-finding techniques have been implemented using observations made either by spinning spacecraft like \textit{Wind} \citep[e.g.,][]{1972Sci...178..743F,1998JGR...10329651R} or three-axis stabilized spacecraft such as \textit{STEREO} or \textit{Cassini} \citep[e.g.,][]{cecconi2008,santolik2003}. The stereoscopic capability of \textit{STEREO}/WAVES \citep{2008SSRv..136..487B} can be used to triangulate the three dimensional position of a radio source at a particular frequency, provided both spacecraft observe the same source quasi-simultaneously. This process can be repeated for different frequencies. This technique has been applied with great success in the past in the study of type III emission \citep{1978JGR....83..616G,2009SoPh..259..255R} but rarely in the study of type II bursts. Several successful campaigns were undertaken using a combination of spacecraft, such as \textit{Helios}, \textit{Ulysses} and \textit{Wind} \citep[e.g.,][]{1976SoPh...48..361B,1995SSRv...72..255R}, demonstrating the success of direction-finding, by mapping the path of accelerated electrons during type III bursts.

In this paper we study the relationship between the interaction of two CMEs and the location of the associated radio sources during an event which occurred on 2010 August 1.  We made use of three space-based instruments with direction-finding capabilities, namely \textit{STEREO}/WAVES experiments \citep{Kaiser08,2008SSRv..136..529B} and \textit{Wind},  along with white-light data from the SECCHI suite onboard \textit{STEREO}.

\section{Observations and Analysis}

\begin{figure}[!h]
\centering
\includegraphics[width=0.89\columnwidth, trim = 60 105 250 55]{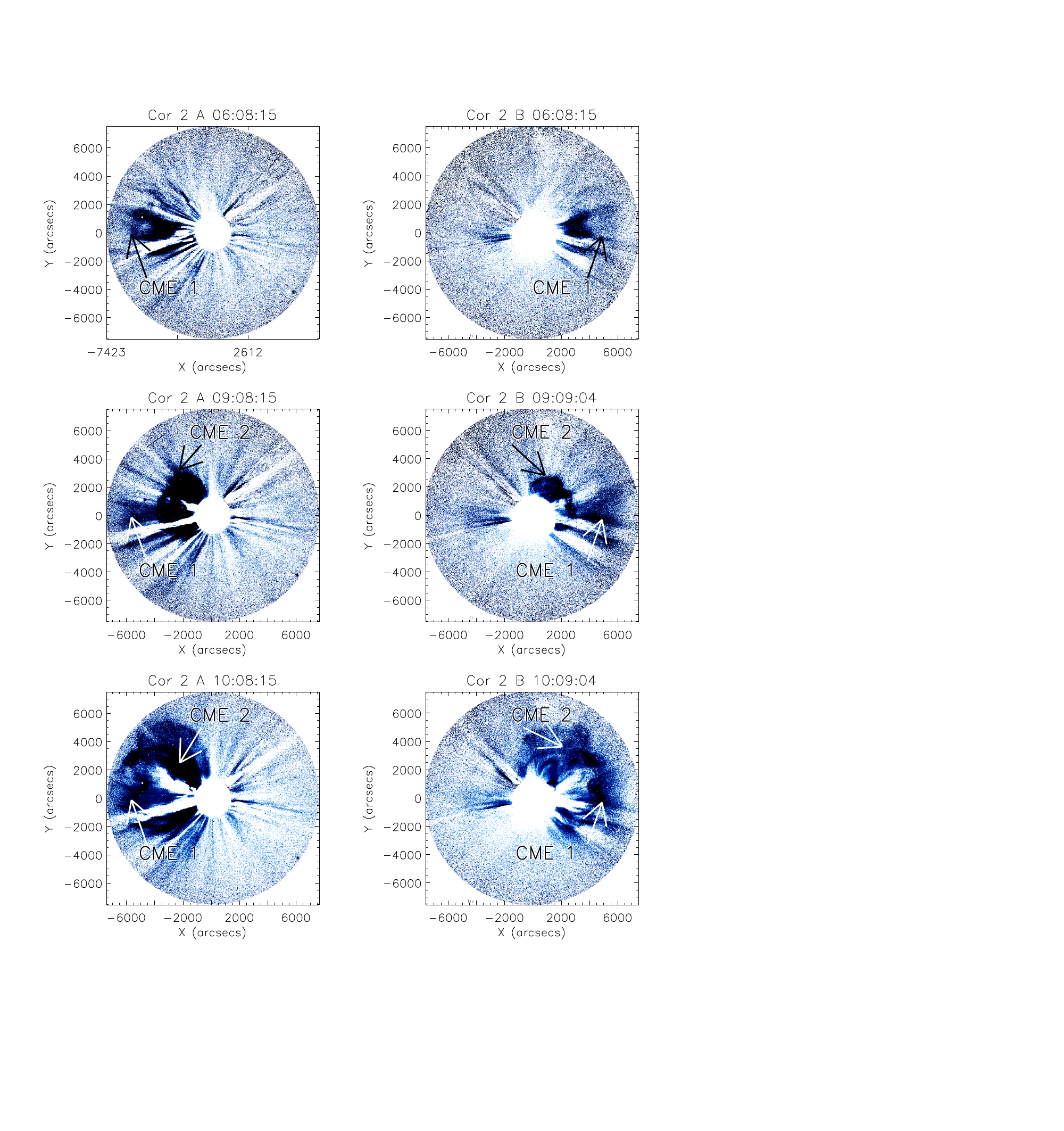}
	\caption{White-light observations of the two CMEs labeled as CME1 and CME2 at three different times during the evolution of the event. The interaction between the CMEs is clearly seen in the bottom frames.}
\label{fig:cmes}
\end{figure}

The period 2010 July 31 to August 2  was characterized by  increased solar activity, exhibiting small flares, filament eruptions and coronal mass ejections \citep{st2011,temmer2011,ying2011}. Of particular interest here is the time during which two CMEs (one slow(CME1), erupted at ~02:00UT and one fast(CME2), erupted at~07:00UT) interacted with each other, resulting in a low frequency type II radio burst observed on 2010 August 1 at about 09:00~UT). The two CMEs in question can be seen in Figure~\ref{fig:cmes} (left panels - Cor2 A, right panels - Cor2 B). Each row corresponds to a time either before (top and middle rows) or during (bottom rows) the interaction period.

\begin{figure}[!h]
\centering
\includegraphics[trim=20mm 0mm 20mm 0mm,width=0.8\columnwidth]{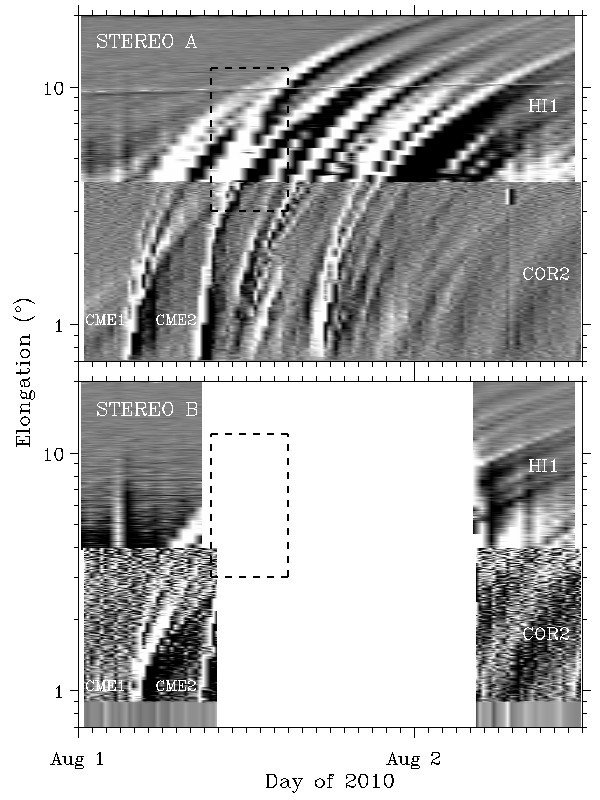}
	\caption{Time\,--\,elongation maps constructed from running-difference images of Cor2 and HI1 along the ecliptic plane for \textit{STEREO} A and B. The dashed box shows the interaction region between the two CMEs labeled CME1 and CME2.}
\label{fig:elongation}
\end{figure}

From these white-light observations, we determined the velocity and time of interaction of the expanding CMEs. An elongation map was constructed from running-difference images of Cor2 and  HI1 along the ecliptic plane of \textit{STEREO}-\textit{A} and -\textit{B}, as described in \citet{2010ApJ...722.1762L}. The filamentary structures in the elongation map (Figure~\ref{fig:elongation}) are the propagating CMEs observed in the period 2010 August 1\,--\,3. Figure~\ref{fig:elongation} shows a fast CME (CME2) that intersects and overtakes a slow CME launched earlier (CME1); the region of interaction is shown by a dashed box. The average speed of the fast CME, derived from Cor2 observations is  $\sim$1138~ $\mathrm{km\,s^{-1}}$  with a liftoff time of $\mathrm{\sim}$07:48~UT from the Sun. The liftoff time of the slow CME was calculated to be 02:48~UT with an average propagation velocity of 730~ $\mathrm{km\,s^{-1}}$ in Cor2 \citep{temmer2011,ying2011}.

Unfortunately, the \textit{STEREO}-\textit{B} Cor1 and HI1 had a data gap of about 18 hours starting at 9:20~UT, which restricted our analysis. It is important to mention that the speed estimates presented above were obtained using an algorithm that fits the distance with a linear model. These results were then compared to the speed derived from adjacent distances with a three-point Lagrangian interpolation, obtaining similar results. From the Lagrangian algorithm, it is possible to derive the error of the computation giving $\pm$ 315~ $\mathrm{km\,s^{-1}}$ and $\pm$ 206~ $\mathrm{km\,s^{-1}}$ for the fast and slow CMEs, respectively. It should be noted that such large errors are present in all methods that measure distances and are not only present in the triangulation method used in this study.

\subsection{Radio emission}
\begin{figure*}[!ht]
\centering
\includegraphics[trim=5mm 10mm 10mm 5mm, width=0.85\textwidth]{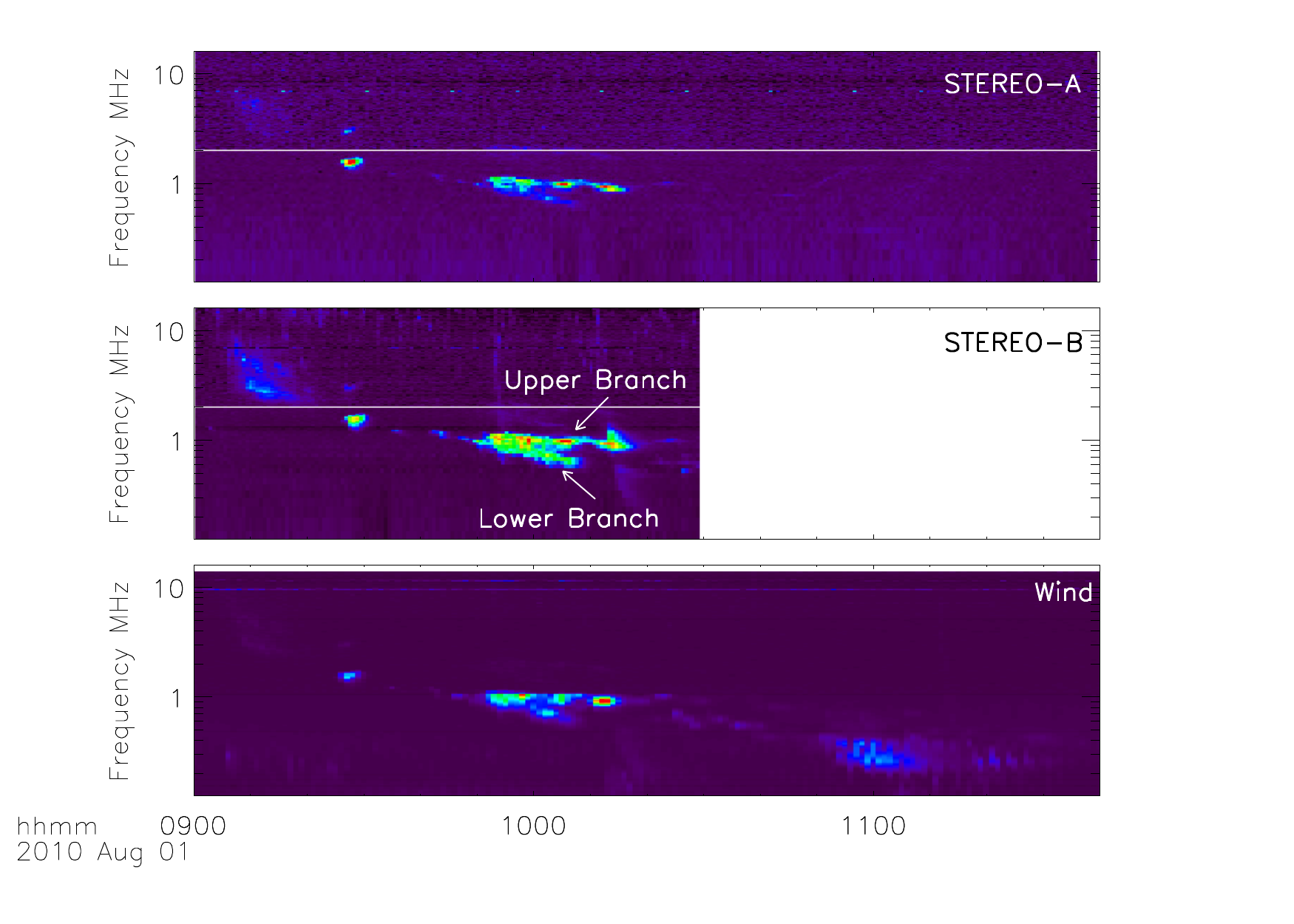}
	\caption{\textit{STEREO}-\textit{A} , \textit{STEREO}-\textit{B} and \textit{Wind} dynamic spectra of the 2010 August 1 type II burst from 09:00~UT to 12:00~UT. The plotted frequency range from 125~kHz to 16.025~MHz for \textit{STEREO} and \textit{Wind}. The color shading represents the intensity of the radio emission measured in arbitrary units.} 
\label{rspec}
\end{figure*}

The  type II radio burst of interest and its modification by CME interaction is shown in Figure~\ref{rspec}. This event was detected simultaneously by both \textit{STEREO} and \textit{Wind} spacecraft. The radio emission was characterized by a slow drifting feature, first observed by \textit{STEREO}-\textit{B} at about 9:10~UT and ending around 11:30~UT. The drift velocity observed by \textit{Wind}/WAVES, starts at 2000~kHz and ends at about 700~kHz.

As Figure~\ref{rspec} shows, the drifting feature in the radio spectra appears to split into two bands at about 9:50~UT. This may be related to the properties of the ambient plasma or, as we show here, the signatures of two interacting CMEs. We fit the \textit{STEREO}/WAVES radio spectra to determine the propagation velocity and the distance at which the emission could be produced, assuming an interplanetary density model \citep{1998SoPh..183..165L}. The radio spectrum was divided in two regions, termed upper and lower branches following the observed division in the spectrogram. Table~\ref{table1} shows the radial distance to the Sun  derived from the model.  From the model the drift velocity was calculated (Table~\ref{table1}), revealing that the lower branch has a higher drift velocity, indicating that this part of the type II radio emission was associated with the fast CME (CME2). The upper branch demonstrated a slower drift velocity which is consistent with original velocity of the first and slower CME, indicating that this was the signature of the slow CME or merged CME front.

\begin{table}[h]
\centering
\begin{tabular}{l}
{
\begin{tabular}{|l|c|c|}
\multicolumn{3}{l}{}\\
\multicolumn{3}{l}{Distance~(AU)}\\
\hline
Branch		&\textit{STEREO}-\textit{A}&\textit{STEREO}-\textit{B}\\
\hline
Lower			&0.025\,--\,0.033		&0.022\,--\,0.036\\
Upper			&0.025\,--\,0.027		&0.025\,--\,0.027\\
\hline
\multicolumn{3}{l}{}\\
\multicolumn{3}{l}{Velocity ($\mathrm{km\,s^{-1}}$)}\\
\hline
Branch		&\textit{STEREO}-\textit{A}&\textit{STEREO}-\textit{B}\\
\hline
Lower  	&1370	&1600	\\
Upper 	&290	&400	\\
\hline
\hline
\end{tabular}
}
\end{tabular}
\caption{Estimated radial distances to the Sun and drift velocities derived from \citet{1998SoPh..183..165L} density model for both branches in the radio spectra and for each \textit{STEREO} spacecraft.}
\label{table1}
\end{table}%

Since the velocities are determined using an interplanetary density model, the results are highly sensitive to any change of the 1 AU electron density (in the model) used for this computation. An uncertainty estimate of 100\,--\,200~$\mathrm{km\,s^{-1}}$ was obtained by calculating the speed using the \citet{1998SoPh..183..165L} density model for variety of ambient electron density values which range from 4 to 7~$\mathrm{cm^{-3}}$ (as observed in in situ data 4 days after the event by \textit{Wind}). These velocities are comparable to the propagation velocities derived from Cor2 observations within the errors of the measurements, suggesting that the split branches observed at 09:50~UT in radio spectra are the signatures of the two interacting CMEs.

\subsubsection{Direction-finding}\label{df}

This event was observed by three spacecraft which, with radio direction-finding capabilities, gives a unique opportunity to study and locate the region or regions responsible for the radio emission in the interplanetary medium. There are different techniques that allow us to determine the distance at which the emission was produced relative to the observer. Some of these make use of electron density models, as demonstrated in the previous section, which provide a direct correlation between the observed frequency and the distance (height) at which they occur \citep[e.g.,][]{1998SoPh..183..165L,2007ApJ...663.1369R}. However, these techniques do not take into account inhomogeneities that may occur in both the interplanetary space and/or the ejected material. Also, the propagation direction cannot be determined by a density model. Other ``direction-finding" techniques, which locate the region of emission by triangulating the position of the radio source at distances of $\mathrm{\sim}$0.1-0.5~AU, have been developed during the last four decades \citep[e.g.,][]{1972Sci...178..743F,cecconi2008,santolik2003,Martinez10}.

\begin{figure}[!h]
\includegraphics[width=0.8\columnwidth, trim = 30 0 100 0]{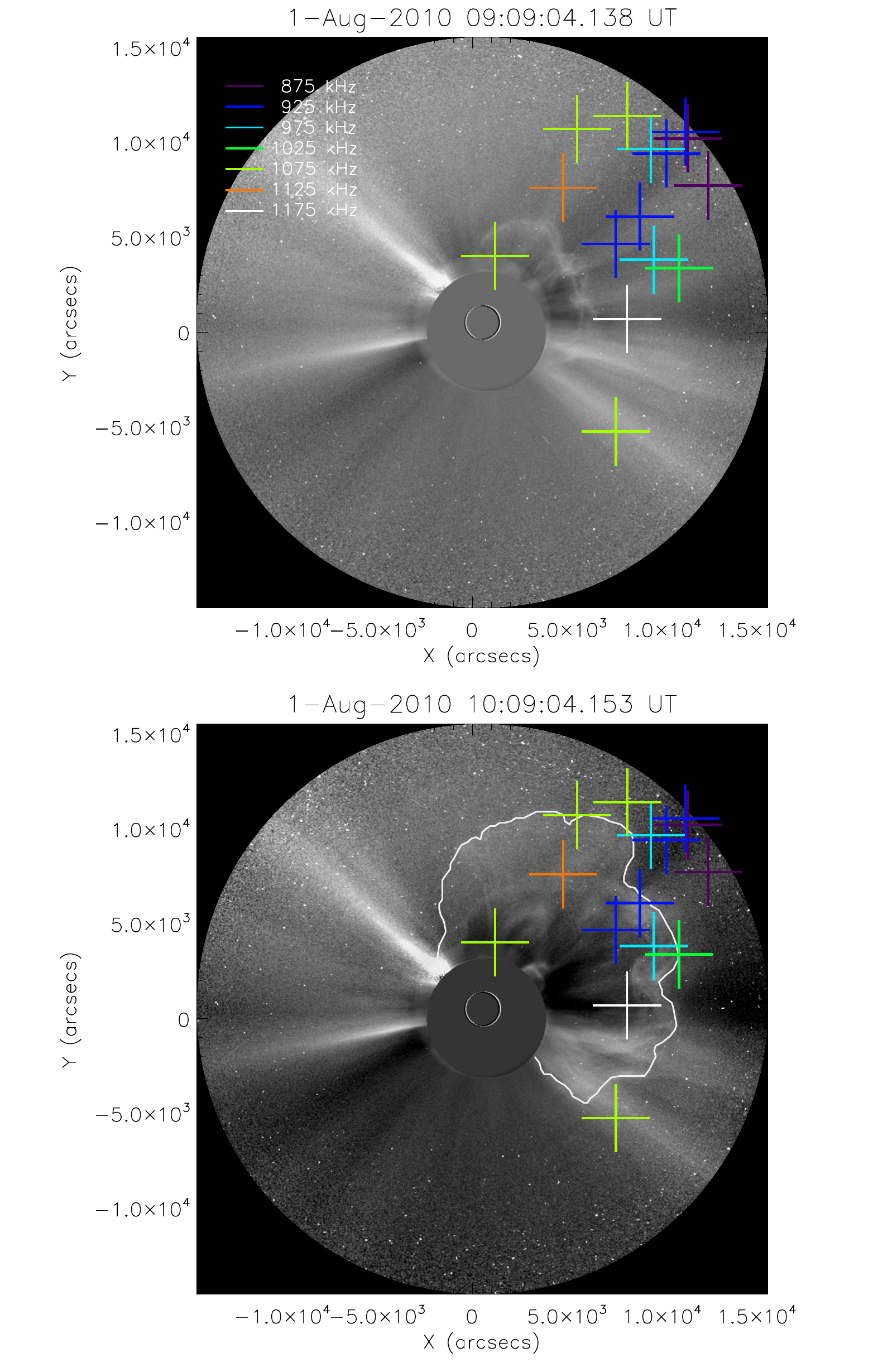}
\caption{\small{\textit{STEREO}-\textit{B} observation of the 2010 August 1 CME from Cor2 at two representative times of type II radio burst, 09:09~UT and 10:09~UT, with line-of-sight direction-finding results from \textit{STEREO}-\textit{B}/WAVES overplotted in color, where color represents different frequencies. The solid white line in lower panel shows the contour of the expanding CME.}}
\label{fig:rp}
\end{figure}

We applied eigenvector and singular value decomposition algorithms  \citep[][respectively]{{Martinez10},{santolik2003}} to determine the arrival direction of radio waves in the frequency range of the High Frequency Receiver 1 instrument onboard \textit{STEREO}. For \textit{Wind}/WAVES data, a modulation technique was applied to retrieve the radio waves' direction of arrival \citep{1972Sci...178..743F} in the range of the Radio Receiver Band 2. The direction of arrival was then characterized by unitary vectors, defined by the azimuths and elevations found by the direction-finding procedure for all  observations in the time range. The spatial positions of the radio sources in interplanetary space were found using a geometrical triangulation algorithm based on \citet{2010ApJ...722.1762L}. For simplicity, we will refer to the combined direction-finding  and triangulation techniques as ``direction-finding''.

\begin{figure*}[htb]
\centering
\includegraphics[width=0.85\textwidth, trim = 70 150 40 50]{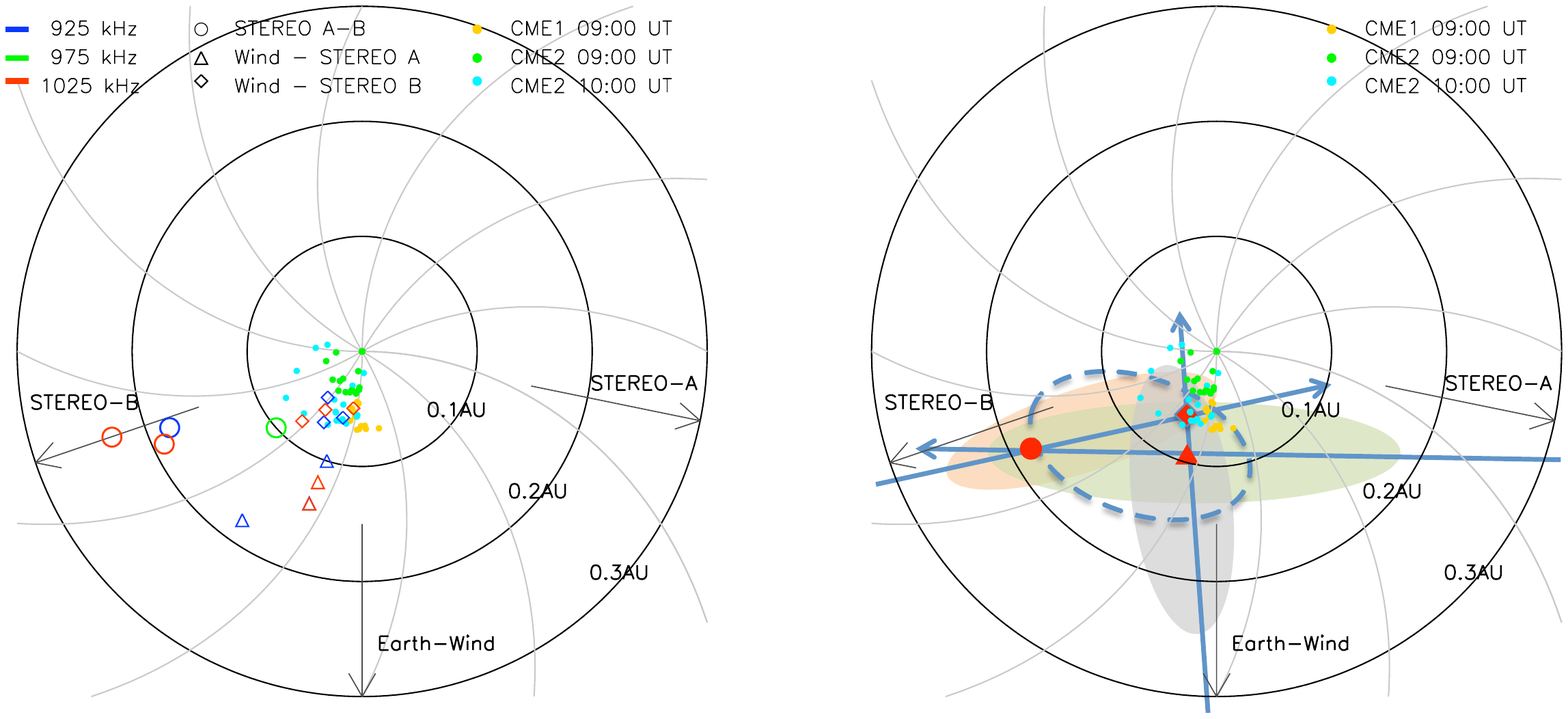}
	\caption{\small{Left: location of the geometrically triangulated positions of the radio sources in interplanetary space for all operational frequencies as seen from the top. The Parker spiral is plotted in gray for context. Right:  radio emission scenario, showing the possible emission region as an extended source propagating between \textit{STEREO}-\textit{B} and \textit{Wind}. The dots represent positions of the two associated coronal mass ejections at different times. The red symbols show the intersection between line-of-sight vectors from the spacecraft represented by the arrows. These are results projected on the ecliptic plane. The overall regions where the direction-Þnding positions are located are represented by the color shaded areas. The dashed ellipse shows the area covering all direction results.}}
\label{fig:pos}
\end{figure*}

In order to determine the location of the type II radio burst relative to the CMEs, white-light images from SECCHI were used. Data from the Cor2 instrument on board \textit{STEREO} B prior to and at the time of the radio burst were used to compare the projected radio direction-finding results (see Figure~\ref{fig:rp}). This type of comparison has been applied before in the study of type II bursts \citep[e.g.,][]{1982AdSpR...2..203W}. The results of our analysis suggest  a close relationship between CME\,--\,CME interaction region and the type II radio burst.  This agrees with the findings of  \citet{2001ApJ...548L..91G} who reported a similar result for a different event.

The dual views of the \textit{STEREO} spacecraft were exploited to identify the location of the radio burst in 3D space, relative to the CME structure. Figure~\ref{fig:pos} shows the front of the CME in heliographic coordinates obtained using the SolarSoft package suite \citep{1998SoPh..182..497F}. The three dimensional locations of two CMEs are shown in Figure~\ref{fig:pos}: the position of the slow CME launched at about 02:00~UT is shown as yellow dots, while the position of the fast CME  launched at approximately 07:00~UT is represented by blue and green dots. Two evolutionary times are shown, 09:00 and 10:00~UT, as during this period the interaction between the two CMEs occurred, with the fast CME overtaking the slow one slightly after 09:00~UT (see Figure~\ref{fig:elongation}). Note that these three-dimensional observations, shown in Figure~\ref{fig:pos}, are projected on the ecliptic plane.

Figure~\ref{fig:pos}~(left) shows the triangulated position of the radio source for the three combinations of \textit{STEREO}-\textit{A}, -\textit{B} and \textit{Wind} spacecraft at the time of the most prominent peaks in the radio flux for three frequencies (925, 975, and 1025~kHz\footnote{For context, the Parker spiral is plotted in Figure~\ref{fig:pos} and was calculated using the formula $\phi = \phi_0 - (\Theta_\sun / V_{\mathrm{sw}} )r$, where $r$, is the radial distance to the Sun, $\phi_0$ is an arbitrary angle, $\theta_\sun$  is the rotational velocity of the Sun ($\mathrm{2~km\,s^{-1}}$), and $V_{\mathrm{sw}}$ is the solar wind velocity ($\mathrm{400~km\,s^{-1}}$).}, see Figure~\ref{peaks}).

The apparent misalignment between the triangulated locations from different spacecraft can be explained by understanding that the triangulation algorithm searches for the position in space where the vectors intersect. This intersection does not necessarily occur at the front of the emitting region, or its  centroid (Figure~\ref{fig:pos}, right frame). Here, it is likely that the source observed is highly extended and complex. Therefore, each spacecraft identified different regions of the extended source (blue dashed ellipse) due to, e.g. the structure of the region and the surrounding local plasma density. Another possible explanation for the apparent discrepancy is that a dense region was located somewhere between the type II radio source and \textit{STEREO}-\textit{A}. This region could scatter the radio waves, leading to an apparent shift of the line-of-sight source position.

\subsubsection{Time-of-flight analysis}

\begin{figure}[!h]
\centering
\includegraphics[width=0.8\columnwidth, trim = 70 50 70 130]{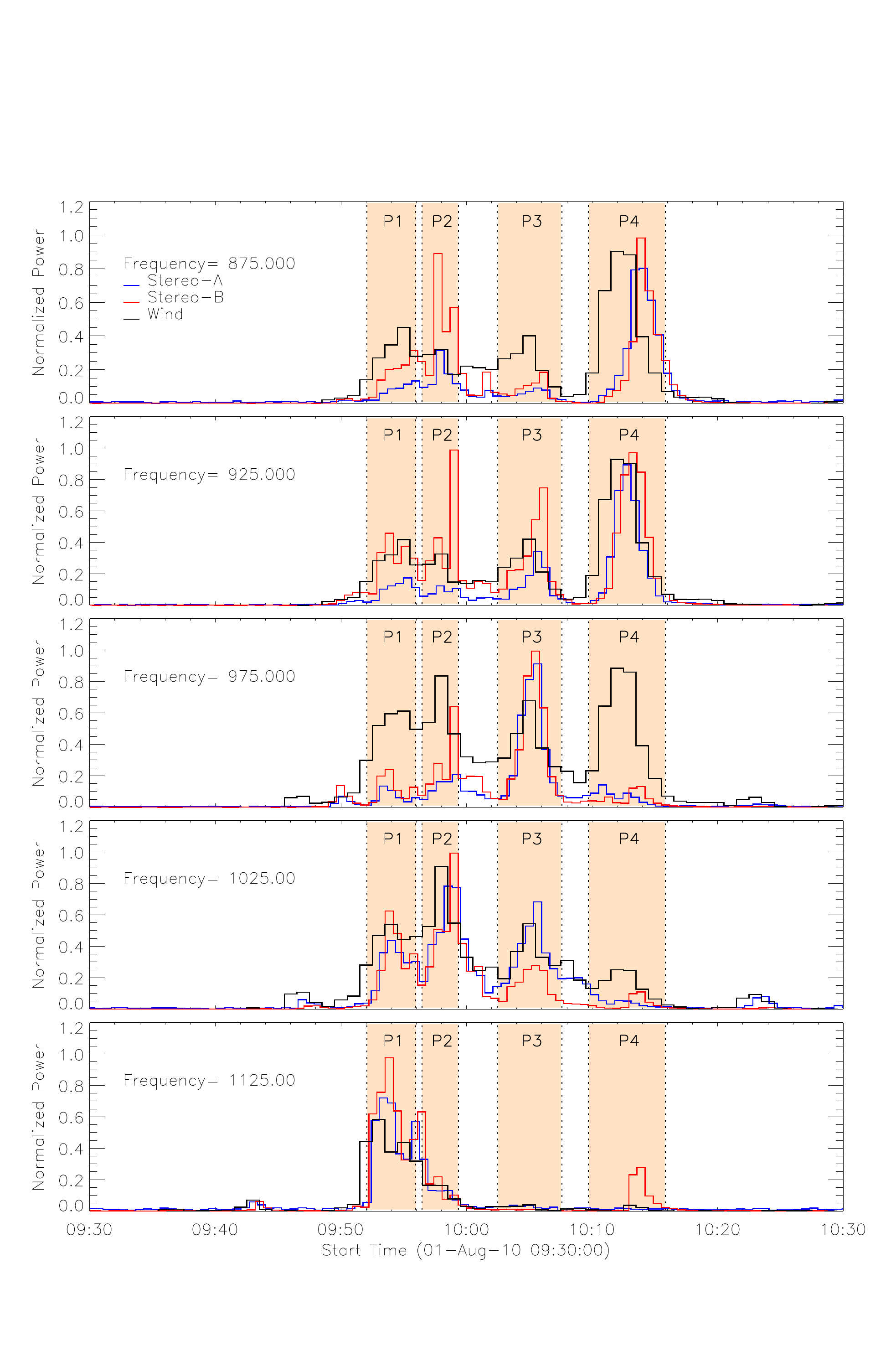}
	\caption{Type II radio burst flux at the observational frequencies. The shaded areas show the most significant peaks of the radio emission. The flux was normalized by the maximum in the observational interval.}
\label{peaks}
\end{figure}

We examined the timing of the radio profiles at the three spacecraft as a control technique to validate the locations and results determined by direction-finding.  We do this by first computing the distance to each spacecraft from the extrapolated locations. Then, times of flight for each spacecraft are computed assuming that the radio emission travels in a straight line from the source centroid to the spacecraft at a constant velocity (the speed of light).  The difference between these two times is compared with the time shift between radio flux profiles at the three spacecraft (see Figure~\ref{peaks}) to determine whether they are consistent with the source locations found using the direction-finding method.  This ``time-of-flight analysis'' assumes that the onsets of the signals at the two spacecraft are the signature of radio emission simultaneously emitted from a single compact source.  The limitations of this ``time-of-flight" analysis are in the temporal resolution of the measurements and errors inherent in the assumptions of compactness and simultaneity.

We find that the time shift (delay) computed from the direction-finding results ranges from $\approx$2 minutes at the lowest frequencies decreasing to $\approx$1 minute at the highest, while the observed delay between the peaks of the emission received by \textit{STEREO}-\textit{A} and -\textit{B} ranges from $\approx$1 to $\approx$0 minute, respectively.  In our analysis we also make use of \textit{Wind} data. Comparing the times of arrivals at the \textit{Wind} and the STEREO spacecraft computed from the direction-finding results, we find an average time delay between $\approx$5.2 minutes and $\approx$1 minute, while the observed time delay between the radio signatures is about $\approx$2 minutes. We found that the direction-finding and the time-of-flight analysis results are consistent within the errors inherent to both techniques. The geometrical configuration suggested by the time-of-flight analysis is consistent to the one obtained by the direction-finding, in which \textit{Wind} is located closer to the radio source that either of the STEREO spacecraft, and also that the radio source is located almost at the same distance from each STEREO spacecraft.

\section{Conclusions}\label{results}

During  2010 July 31\,--\,August 2 a series of CMEs and their associated type II and type III radio bursts were observed. In particular, an interplanetary type II burst was detected by instruments onboard the STEREO and \textit{Wind} spacecraft on 2010 August 1 at about 09:00~UT. The close timing between the type II radio burst and the interaction of two coronal mass ejections suggests that the radio emission is a consequence of this interaction. A similar event was analyzed previously by  \citet{2001ApJ...548L..91G},  who concluded that the interaction between slow CME and a fast one resulted in the enhancement of the radio emission during the transit of the fast CME shock front through the core of the slow CME. 

Using white-light and radio observations we estimated the propagation velocities of the two CMEs. We found that the velocities derived from radio observations are comparable to the propagation velocities derived from coronagraph observations. This suggests that the branches in the radio spectra, observed at 09:50~UT, are the signatures of the two interacting CMEs. Using the density model of \citet{1998SoPh..183..165L} we also estimated the distance at which the radio emission was produced, was between 0.025 and 0.043~AU. This is in agreement with the radio direction-finding results, which give a distance about 0.01\,--\,0.05~AU. From white-light observations, we determined that the shock front  propagated $\sim$20$^\circ$ east of the Sun\,--\,Earth line (i.e. between \textit{STEREO}-\textit{B} and \textit{Wind}), which is about the same angular separation derived by the direction-finding technique. The obtained propagation direction is in agreement with finding of \citet{temmer2011} and \citet{ying2011}.

We successfully applied three radio direction-finding techniques \citep{1972Sci...178..743F, santolik2003,Martinez10} to the 2010 August 1 type II radio burst and determined the direction of arrival of the radio emission. The data analysis shows that the radio sources locations are spread over a large area covering about 4$\mathrm{^\circ}$, suggesting that the radio source has an extended and complex structure in nature, perhaps composed of multiple radio emitting regions which may have a common origin. We found good consistency between the triangulated white-light positions and the \textit{Wind}\,--\,\textit{STEREO}-\textit{B} triangulated positions. Using STEREO-A, we found a discrepancy that can be explained by the complexity of the source and the surrounding material. Since neither the emitting region nor the medium are homogenous, it is possible that the radio source was partially occulted in the direction of STEREO-A by a dense solar wind region. This may explain the relatively low power observed in the STEREO-A spectrogram and can also account for scattering of radio waves, which consequently will shift the apparent position of the radio source.

By comparing these positions with white-light features in the STEREO coronagraph data and their derived positions as described in Section~\ref{df}, we found that the radio emission is the result of the interaction between two expanding CMEs. Figure~\ref{fig:rp} shows that the positions derived from the direction-finding match the features observed in the coronagraph images, suggesting the relation between the type II radio emission and the interaction region of two expanding CMEs.
 
Radio direction-finding has proven to be a powerful technique in the study of CMEs and associated type II radio bursts. By using these techniques, it is possible to determine the heliographic distance of a radio source, which is independent of any density model.  In general, the limitation of these techniques is given by the frequency of observations and the properties of the radio emission region. In the case of metric wavelengths this error can be about 1$^\circ$ in azimuth and elevation. Nevertheless,our study shows that with good observations, the techniques give results that can be directly compared with observations at other wavelengths and show the likely emission region. The application of radio direction-finding methods to data acquired by future missions, such as Solar Probe Plus and Solar Orbiter, will prove to be crucial in our understanding of CMEs and type II radio bursts.

\bibliographystyle{chicago}

\end{document}